\documentclass[fleqn,10pt]{wlscirep}
\usepackage[utf8]{inputenc}
\usepackage[T1]{fontenc}

\usepackage{physics}
\usepackage{siunitx}
\usepackage{subcaption}

\title{Demonstration of a Compatibility-Based Childcare Support Service using Quantum Annealing}

\author[1,*]{Yuuma Matsumoto}
\author[2]{Taisei Takabayashi}
\author[2,3]{Rima Sato}
\author[2]{Rumiko Honda}
\author[1,2,4,5]{Masayuki Ohzeki}
\affil[1]{Department. of Physics, Institute of Science Tokyo}
\affil[2]{Graduate School of Information Sciences, Tohoku University}
\affil[3]{MUSASI D{\&}T Co., Ltd.}
\affil[4]{Research and Education Institute for Semiconductors and Informatics, Kumamoto University}
\affil[5]{Sigma-I Co., Ltd.}

\affil[*]{matsumoto.y.f4be@m.isct.ac.jp}

\begin{abstract}
In contemporary Japan, isolated parenting has become a serious social issue, increasing psychological stress on parents and potentially affecting children’s development. 
Existing childcare support services tend to focus on physical assistance, while psychological support and community connections remain insufficient.
To address this gap, we developed a service that connects parents with senior community members who have parenting experience, aiming to provide psychological support and foster intergenerational exchange. 
Achieving high-quality matching requires considering pair compatibility, balancing supporter workload, and handling scheduling constraints, which can be formulated as a combinatorial optimization problem.
We designed a matching framework using the Quadratic Unconstrained Binary Optimization (QUBO) formulation and evaluated quantum annealing (QA) against simulated annealing (SA). 
QA achieved higher solution quality and diversity, particularly for larger problem instances. 
Furthermore, a proof-of-concept field experiment conducted in Sendai City, Japan, demonstrated that the framework can generate multiple high-quality matching candidates, enabling flexible scheduling in real-world operations.
\end{abstract}
\begin{document}

\flushbottom
\maketitle
\thispagestyle{empty}

\section*{Introduction}
In contemporary Japan, where the trend toward nuclear families and the weakening of ties with local communities continue to advance,
an increasing number of parents experience anxiety and a sense of isolation in childrearing.
Recent surveys conducted by the Cabinet Office have revealed that, compared to 2015, the proportion of parents who report feeling "psychological fatigue" or "physical fatigue"
associated with childcare has increased, highlighting that parenting is being perceived as a heavier burden than in the past\cite{naikaku}.
Furthermore, a study targeting parents of children aged three to five in Japan demonstrated that support related to childcare from partners, grandparents, and non-relatives 
does not exert a direct effect on children's social development. Instead, such support influences children's outcomes indirectly, by shaping parents' psychological states and parenting styles\cite{morita2021childcare}.
This finding suggests that reducing mothers' feelings of anxiety and loneliness regarding childrearing may contribute to fostering children's development.
At present, many childcare support services primarily focus on alleviating physical burdens through means such as temporary childcare services and housework assistance.
However, it is difficult to argue that sufficient support is being provided to address the more diffuse and persistent feelings of anxiety and social isolation that parents often experience during childrearing.

Building upon this background, this study proposes a compatibility-oriented matching support model aimed at providing psychological reassurance to parents raising children (referred to as "users")
by connecting them with senior members of the local community who have prior parenting experience (referred to as "supporters").
In particular, subjective factors such as personality traits and parenting beliefs are quantified through questionnaires and incorporated into the model as compatibility scores between users and supporters.
We assume that, following the determination of the user-supporter pairing, a single home-visit interaction will be conducted within a limited timeframe.
This design is intended both to reduce the psychological and logistical barriers for participants and to enable the verification of the short-term effects of such psychosocial support.
Consequently, the matching process must take into account practical constraints, such as the availability of overlapping visit schedules, to ensure that interactions can be successfully carried out.
A further consideration is the potential for a disproportionate concentration of matches on particular supporters, which could lead to an uneven distribution of burden.
This risk is especially pronounced when certain supporters achieve high compatibility scores with many users, and allowing unlimited matches in such cases would place excessive strain on those individuals.
To avoid this outcome, we introduce a constraint limiting the number of matches each supporter can undertake.
The matching framework is thus designed to balance the maximization of overall compatibility scores with the equitable distribution of supporter workload.
To simultaneously satisfy these multiple practical requirements while deriving an optimal matching configuration, the problem is formulated as a combinatorial optimization problem
using the Quadratic Unconstrained Binary Optimization (QUBO) formulation.

The QUBO formulation is a widely used optimization model in the field of quantum computing and has attracted attention as a unified framework 
capable of representing a wide variety of combinatorial optimization problems\cite{Glover2022}.
In recent years, Quantum Annealing (QA), an algorithm based on the principles of quantum mechanics, has gained increasing attention as a method for solving QUBO problems\cite{kadowaki1998}.
QA searches for the ground state of an Ising model, and because many combinatorial optimization problems can be transformed into the Ising model\cite{lucas2014ising},
its applications have rapidly expanded across a wide range of domains, including traffic signal control\cite{neukart2017trafficflowoptimizationusing,Inoue2021,shikanai2023},
optimization of manufacturing processes\cite{ohzeki2019control,Haba2022}, financial portfolio construction\cite{rosenberg2016solving,Venturelli2019},
scheduling in steel manufacturing\cite{yonaga2022quantum},
and optimization of machine learning algorithms\cite{PhysRevX.8.021050,OMalley2018,Sato2021,Urushibata_2022,hasegawa2023kernellearningquantumannealer,Goto_2025}.
Compared with classical QUBO solvers, QA has the potential to explore the solution space more broadly and escape from local minima by leveraging quantum tunneling effects. 
This capability is particularly promising for large-scale or highly rugged optimization landscapes, where conventional solvers often struggle to find diverse high-quality solutions.

The structure of this paper is as follows. In the next chapter, we describe the formulation of the matching problem between users and supporters,
including the design of compatibility scores, the treatment of constraints, and the parameter tuning procedures for the QUBO solver.
In Chapter 3, we analyze the results of solving random QUBO instances and demonstrate that, for relatively large-scale problems,
QA is capable of sampling a broader diversity of high-quality matching solutions.
We subsequently present the outcomes of a proof-of-concept field experiment conducted in Sendai City, Japan,
which validates the effectiveness and practical applicability of the proposed matching model.
Finally, Chapter 4 discusses the implications of the results and outlines future research directions.

\section*{Methods}
\label{sec:methods}
This chapter describes the mathematical formulation of the optimization problem for the matching problem between users and supporters.
This study aims to implement a matching strategy that maximizes the total compatibility scores computed for each user–supporter pair. 
These compatibility scores are scalar values quantified based on information obtained through questionnaires, which capture traits such as personality, values, and attitudes toward childcare and household tasks. 
They serve as indicators of psychological affinity between pairs.

Since the proposed matching model assumes that only one visit will be conducted within a specified period, it is also necessary to consider practical constraints, such as the time slots during which matching activities can be carried out and the maximum number of users that can be assigned to each supporter. Some of these constraints are incorporated into the QUBO formulation as penalty terms, while others are handled by pre-filtering to narrow the feasible candidate pairs.

\subsection*{Questionnaire design and compatibility quantification}
Compatibility between users and supporters is quantitatively evaluated based on questionnaire responses that capture traits and values.
The questionnaire comprises 10 items grouped into the following categories to provide a multifaceted profile:

\begin{itemize}
    \item Cognitive and personality tendencies (three items)
    \item Values related to daily life and household tasks (two items)
    \item Communication style and interpersonal preferences (three items)
    \item Attitudes toward learning and interests (two items)
\end{itemize}

\begin{figure}[htbp]
  \centering
  \includegraphics[width=0.8\textwidth]{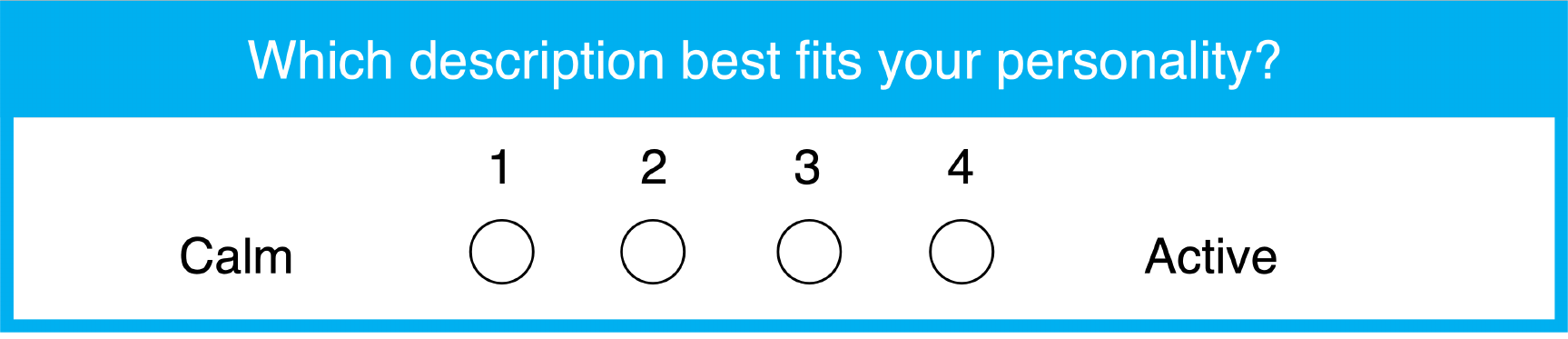}
  \caption{
    Schematic diagram of survey items. 
    This diagram illustrates example items related to personality.
    Respondents are asked to choose the option that best reflects their tendency from four available choices.
  }
  \label{fig:questionaire_example}
\end{figure}

Figure~\ref{fig:questionaire_example} shows an example questionnaire item related to personality,
in which respondents are asked to indicate whether their personality is closer to "Calm" or "Active".
Each item provides four levels of response options, and participants are instructed to select the option that most closely reflects their own tendency.

Based on the questionnaire responses, an item score $m_{ij}^{\mu}$ is calculated for each item $\mu$ for the pair consisting of user $i$ and supporter $j$,
thereby quantifying their compatibility on that particular item.
For example, in the case illustrated in Figure~\ref{fig:questionaire_example}, the item score $m_{ij}^{\mu}$ is defined according to the degree of agreement between the user's and supporter's responses for that item, as follows:
\begin{equation}
  m_{ij}^{\mu} = 
  \begin{cases}
    3 & \quad \text{(exact match)} \\
    2 & \quad \text{(1-point difference)} \\
    1 & \quad \text{(2-point difference)} \\
    0 & \quad \text{(3-point difference)} \\
  \end{cases}
\end{equation}
This formulation assumes that pairs consisting of two calm or active individuals are judged to be more compatible.
Such numerical indicators are defined for each item, and ultimately, all the item scores are summed to compute the overall compatibility score $M_{ij}^{\mu}$ for the pair of user $i$ and supporter $j$:
\begin{equation}
  M_{ij} = \sum_{\mu} m_{ij}^{\mu}\;.
\label{eq:compatibility_score}
\end{equation}

\begin{figure}[t]
  \centering
  \includegraphics[width=0.9\textwidth]{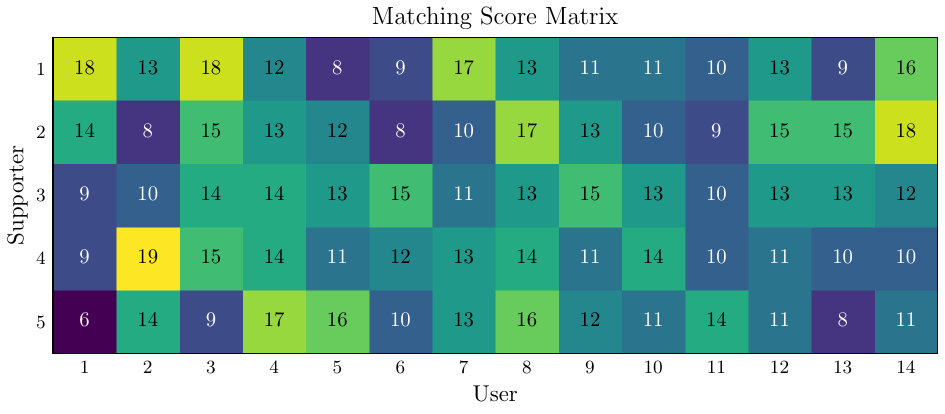}
  \caption{
      Visualization of the overall compatibility score $M_{ij}$ calculated using random data of 14 users and five supporters.
      The combination of users and supporters corresponds to each cell in the matrix, and the number inside represents the compatibility score $M_{ij}$.
  }
  \label{fig:matching_matrix_dummy}
\end{figure}
Figure~\ref{fig:matching_matrix_dummy} shows an example of the compatibility scores $M_{ij}$ calculated using random data for 14 users and five supporters.
For instance, user 1 has a high compatibility score with supporter 1, making this a pair we would ideally like to realize as a matching outcome.
In contrast, the compatibility score between user 1 and supporter 5 is low, and such a pairing should be avoided in the matching process.

\subsection*{Constraints for realistic matching}
\begin{figure}[htbp]
  \centering
  \includegraphics[width=0.9\textwidth]{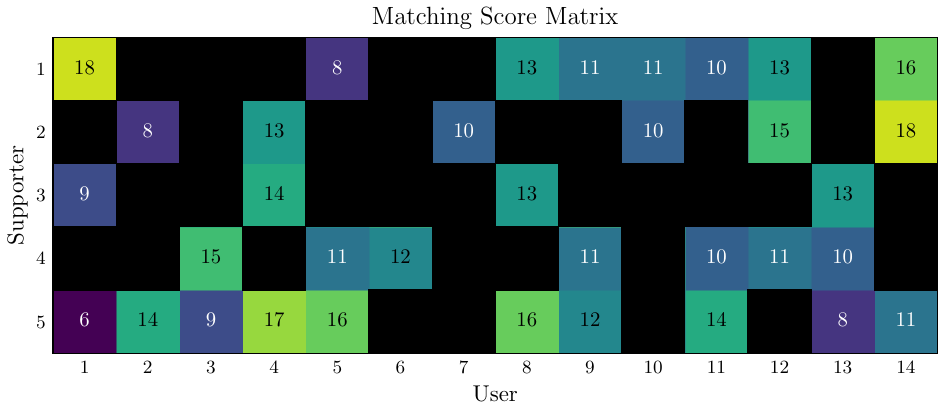}
  \caption{
    The masked version of Figure \ref{fig:matching_matrix_dummy} shows pairs excluded from matching candidates by filtering.
  }
  \label{fig:matching_matrix_dummy_masked}
\end{figure}
In the actual service design, it is necessary to consider maximizing the compatibility scores between users and supporters and practical operational constraints in the service delivery.
The primary constraints to be considered in the matching process are as follows:
(1) the number of users assigned to each supporter,
(2) the number of supporters assigned to each user,
(3) time constraints regarding the availability of users and supporters, and
(4) conditions regarding whether a supporter can provide care for infants.
These constraints can be broadly classified into two categories.
The first category, which includes constraints (1) and (2), can be incorporated into the QUBO objective function as penalty terms.
These constraints can be expressed algebraically based on the number of assignments per user or supporter. They can thus be naturally represented as quadratic penalties within the binary optimization framework.
For example, if the number of users assigned to a supporter deviates from the required number, a term proportional to the square of the difference between the actual and target number of assignments can be added to suppress such violations.
These terms are well-suited for the QUBO framework, which expresses objectives and constraints using binary variables and quadratic interactions.

On the other hand, constraints such as (3) and (4) are addressed by pre-filtering infeasible user-supporter pairs before the optimization.
These constraints typically involve complex logical conditions or continuous variables, making implementing them as algebraic penalties on binary variables challenging.
Directly incorporating such constraints into the QUBO could significantly increase the model's complexity and require large penalty weights, potentially degrading optimization performance.
Moreover, excluding infeasible pairs via pre-filtering also helps reduce the number of qubits required for the optimization.

Figure~\ref{fig:matching_matrix_dummy_masked} shows an example in which the original compatibility score matrix in Figure~\ref{fig:matching_matrix_dummy} has been masked to reflect pairs eliminated by pre-filtering. 
The cells corresponding to excluded pairs are blacked out. They represent pairs that violate constraints (3) or (4) and, therefore, have no matching feasibility.

It is important to note that the objective and computational complexity of pre-filtering and exhaustive search are fundamentally different.
For example, when there are $n$ users and $m$ supporters, finding the optimal assignment of each user to a supporter through exhaustive search would require enumerating $m^n$ possible combinations.
In contrast, pre-filtering independently checks the constraints for each user-supporter pair and eliminates infeasible ones.
Even if all possible pairs are considered, the overall computational complexity remains on the order of $mn$.
Thus, while exhaustive search is computationally impractical for large problem sizes, pre-filtering remains a feasible preprocessing step. 
Consequently, efficiently extracting only the feasible pairs through pre-filtering and subsequently optimizing those pairs is a practical approach that can be realistically implemented in actual operations.

\subsection*{Naive formulation}
Let $U$ denote the set of users and $S$ the set of supporters. The problem of determining pairs based on compatibility scores can be formulated as a matching problem on a bipartite graph $G=(U\cup S, E)$.
Here, $E\subset U\times S$ is the set of pairs that remain feasible after pre-filtering and corresponds to the edges of the bipartite graph.
In other words, if a user $i$ and a supporter $j$ are connected by an edge $e$, we write $i=t(e)$ and $j=h(e)$. Each edge $e \in E$ is associated with a compatibility score $M_e \in \mathbb{R}_{\geq 0}$.

The objective of the matching is to determine user–supporter pairs from $E$.
For this purpose, we define the decision variable $x_e$ using a one-hot encoding as follows:
\begin{equation}
  x_e = 
  \begin{cases}
    1 & \text{if user } i=t(e) \text{ is matched with supporter } j=h(e), \\
    0 & \text{otherwise}.
  \end{cases}
\end{equation}
Using this decision variable, the matching problem of maximizing the total compatibility score can be formulated as:
\begin{align}
\max_x \quad & \sum_{e\in E} M_{e} x_{e}, \\
\text{s.t.} \quad 
& \sum_{e\in E: t(e)=i} x_{e} = 1, \quad \forall i\in U, \label{eq:constraints_for_user}\\
\text{s.t.} \quad
& \sum_{e\in E: h(e)=j} x_{e} = C_j, \quad \forall j\in S. \label{eq:constraints_for_supporter}
\end{align}
Constraint~\eqref{eq:constraints_for_user} ensures that each user is matched with exactly one supporter,
while Constraint~\eqref{eq:constraints_for_supporter} ensures that each supporter is assigned exactly $C_j$ users.
For instance, if there are $n$ users and $m$ supporters and we wish to distribute the workload among supporters evenly, each supporter is assigned $C_j=n/m$ users.
By incorporating these constraints as penalty terms into the objective function, the problem can be transformed into the following QUBO formulation:
\begin{equation}
  \min_x \quad
    - \sum_{e\in E} M_{e} x_{e}
    + \lambda_1 \sum_{i\in U} \left( \sum_{e\in E: t(e)=i} x_{e} - 1 \right)^2
    + \lambda_2 \sum_{j\in S} \left( \sum_{e\in E: h(e)=j} x_{e} - C_j \right)^2
    . \label{eq:qubo_objective_ideal}
\end{equation}
where $\lambda_1$ and $\lambda_2$ are hyperparameters that control the strictness of the constraints. Solving this QUBO yields a matching that maximizes the overall compatibility score while satisfying the given constraints.

However, this formulation requires $|E|$ decision variables. Since $|E|$ depends on the combinations of users and supporters, it grows exponentially as the problem size increases, causing the size of the QUBO to expand rapidly.
In addition, the constraints in Equation~\eqref{eq:qubo_objective_ideal} make it difficult for solvers such as QA and SA to perform well, because the penalty terms generate a higher barrier between states.
Previous studies relaxed the original formulation based on the concepts of statistical mechanics and operations research to mitigate the difficulty of solving the optimization problem with such constraints beyond the penalty method, which is a standard method\cite{doi:10.7566/JPSJ.94.054003,yonaga2020solvinginequalityconstrainedbinaryoptimization,KoukiYonaga2022ISIJINT-2022-019,Ohzeki2020,doi:10.7566/JPSJ.92.113002}.
In this study, we utilize another way to avoid performance degradation: focusing on the properties of the matching problems.
A key observation in practical matching scenarios is that the actual matches tend to focus on pairs with relatively high compatibility.
As a result, pairs with very low compatibility are unlikely to be selected in optimal or near-optimal solutions.
This insight suggests that excluding such unlikely pairs does not substantially affect the matching results.
In practice, this means that most decision variables corresponding to low-compatibility pairs take $x_e=0$, creating redundancy if every possible pair is retained in the naive formulation.
By taking advantage of this property, we next introduce an approximate formulation that reduces the problem size and makes it possible to handle large-scale matching problems more efficiently.

\subsection*{Approximate formulation}
By embedding the matching information directly into the binary value of each decision variable, the problem size can be maximally compressed. 
For example, suppose we impose the constraint that each user must be matched with one of the top two supporters regarding compatibility score, and store information on these two supporters for each user. In that case, it becomes possible to represent the user’s matching with a single binary variable.
This idea is based on the formulation proposed in the context of evacuation optimization, whose effectiveness has also been demonstrated in optimizing connections between mobile phones and base stations \cite{ohzeki2023qubits,takabayashi2025optimization}, making it suitable for application to the current matching problem as well. 
In evacuation optimization, evacuees must be assigned to shelters by considering distance and shelter capacity constraints. 
Similarly, mobile phone networks aim to determine connection patterns that maximize communication quality while balancing resource usage across base stations.
These problem settings are structurally similar to the user–supporter matching problem. 
The compatibility score between user $i$ and supporter $j$ corresponds to the distance between an evacuee and a shelter or the communication quality between a mobile phone and a base station.
Likewise, the number of users $C_j$ assigned to each supporter plays a role analogous to the capacity limit of a shelter or the connection limit of a base station.

Based on this idea, we define the decision variable as follows:
\begin{equation}
  x_i = 
  \begin{cases}
    1 & \text{if user } i \text{ is matched with supporter having the highest compatibility score,}\\
    0 & \text{if user } i \text{ is matched with supporter having the second highest compatibility score}.
  \end{cases}
\end{equation}
In this definition, each user is matched only with one of their top two supporters regarding compatibility.
This allows a second-best option to be selected when the most compatible supporter has already been assigned to many other users, thereby maintaining overall balance. 
Owing to the definition of the decision variable, the user-side matching constraint Equation~\eqref{eq:constraints_for_user} is inherently satisfied.
As a result, this approximate formulation contains fewer penalty terms and constraints than the exact formulation~\eqref{eq:qubo_objective_ideal}, which reduces the effort required for tuning penalty parameters and facilitates a more straightforward optimization process in practice.
The formulation is as follows:
\begin{align}
  \max_x \quad & \sum_{i\in U} \left( M^{(1)}_{i}x_i + M^{(2)}_{i}(1-x_i) \right), \\
  \text{s.t.} \quad 
  & \sum_{i\in U_1(j)} x_i + \sum_{i\in U_2(j)}(1-x_i) = C_j, \quad \forall j\in S. \label{eq:constraints_for_supporter_approx}
\end{align}
Here, $M_i^{(1)}$ and $M_i^{(2)}$ denote the first and second highest compatibility scores for user $i$, respectively.
The sets $U_1(j)$ and $U_2(j)$ represent the sets of users for whom supporter $j$ is ranked first and second, respectively.
The QUBO formulation of this problem is expressed as:
\begin{equation}
    \min_x \quad
    - \sum_{i\in U} \left(M^{(1)}_{i}x_i + M^{(2)}_{i}(1-x_i)\right)
    + \lambda \sum_{j\in S} \left( \sum_{i\in U_1(j)}x_i + \sum_{i\in U_2(j)}(1-x_i) - C_j \right)^2
    . \label{eq:qubo_objective_approx}
\end{equation}
where $\lambda$ is a hyperparameter that controls the strictness of the supporter-side constraints.

In this formulation, the number of variables equals the number of users $|U|$, representing a significant compression compared to the naive formulation~\eqref{eq:qubo_objective_ideal}.
Moreover, the solution quality is expected to remain high because each user can only be matched with their top two supporters regarding compatibility score.
This reduction is particularly beneficial for QA, where the limited number of available qubits often becomes a bottleneck.
This approach reduces the number of variables, making it possible to obtain practical approximate solutions while using fewer qubits.
Similarly, in the corresponding classical method, Simulated Annealing (SA) \cite{SA}, the reduction in the number of variables can be expected to improve computational efficiency.
However, this formulation is still an approximation of Equation~\eqref{eq:qubo_objective_ideal} and does not necessarily satisfy the supporter assignment constraint~\eqref{eq:constraints_for_supporter} exactly.
As a result, in the obtained matching, it is possible that some supporters may be assigned more users than their capacity allows or fewer users than required.
Therefore, there is an inherent trade-off between strictly satisfying the supporter assignment constraints~\eqref{eq:constraints_for_supporter} and reducing the number of decision variables to improve computational efficiency. 
Depending on the application, selecting a formulation that appropriately balances accuracy and efficiency is necessary.

\subsection*{Parameter tuning and solver settings}
In the next Result chapter, we compare the solutions obtained using QA and SA for the exact matching formulation given by Equation~\eqref{eq:qubo_objective_ideal}. 
Since both SA and QA are stochastic optimization methods, appropriately tuning hyperparameters according to the problem structure is essential to obtain high-quality solutions.

The first key hyperparameters to be tuned are the penalty coefficients $\lambda_1$ and $\lambda_2$ in Equation~\eqref{eq:qubo_objective_ideal}.
These coefficients determine the weight of the penalty for constraint violations: the larger their values, the higher the energy assigned to solutions that violate the constraints.
Consequently, setting $\lambda_1$ and $\lambda_2$ to large values makes it more likely that the low-energy states of the QUBO will correspond to feasible solutions satisfying the constraints.
However, suppose these coefficients are set excessively large. In that case, the energy of the constraint terms will dominate the energy landscape, effectively prioritizing the avoidance of constraint violations over the minimization of the objective function. 
As a result, the solver may become trapped in local optima that satisfy the constraints but fail to minimize the objective function sufficiently. 
Moreover, when the penalties are overly strong, it becomes more difficult for the solver to temporarily relax constraints and explore solutions with better objective values, limiting the exploration of the solution space within a finite time.
This restricts the inherent exploration capability of stochastic methods such as QA and SA and can hinder convergence to high-quality optimal solutions.
Therefore, the penalty coefficients $\lambda_1$ and $\lambda_2$ must be carefully adjusted to balance constraint enforcement and the ability to explore the solution space. 
In this study, considering the symmetry of the formulation, we set $\lambda_1=\lambda_2$. To select an appropriate value, we conducted searches using SA on several representative problem instances,
evaluated the quality of the resulting solutions, and then applied the same value to the QA configuration.

For the execution of SA, we used the \texttt{SASampler} provided by OpenJij \cite{SASampler}.
In SA, the granularity of the search can be controlled by adjusting the lower and upper bounds of the inverse temperature schedule, $\beta_{\text{min}}$ and $\beta_{\text{max}}$.
In general, a low inverse temperature allows for a broad exploration of the solution space, while a high inverse temperature focuses on local optimization.
In the default configuration of \texttt{SASampler}, a different inverse temperature schedule is automatically assigned to each problem instance. 
However, this setting resulted in variability in the quality and reproducibility of the solutions obtained. 
Therefore, in this study, we adopted a common inverse temperature schedule of $(\beta_{\text{min}}, \beta_{\text{max}}) = (0.02, 2.0)$ for all instances to ensure the stability and reproducibility of the algorithm.
Since SA is a stochastic method, statistical variability arises in the solutions obtained from each run. 
To ensure reproducibility and reliability, we set the number of samples to $\texttt{num\_reads}=1000$.
Additionally, the parameter \texttt{num\_sweeps} controls the number of Monte Carlo steps in the Markov Chain Monte Carlo process used to generate each sample, indicating how many spin updates are performed in a single sampling.
In this study, we adopted the default setting of \texttt{SASampler}, $\texttt{num\_sweeps}=1000$.

For the execution of QA, we used \texttt{Advantage\_system6.4} by D-Wave Systems. 
To solve a QUBO problem on hardware, logical variables must be appropriately mapped onto the physical structure of qubits, a process known as embedding. 
Due to the physical connectivity constraints among qubits in quantum annealers, it is often not possible to map the couplings between variables in the logical graph onto physical qubits one-to-one.
To overcome this limitation, a single logical variable can be split across multiple physical qubits, which are then connected by strong couplings to behave as a single variable. 
The set of physical qubits configured in this way is called a chain, and logical consistency is maintained by ensuring that all qubits in the chain take the same spin state.
In this study, we used \texttt{minorminer}, included in the D-Wave Ocean SDK. We applied the minor-miner technique \cite{cai2014practicalheuristicfindinggraph} to map the logical graph onto D-Wave's physical graph, the Pegasus topology.

To ensure the stable operation of chains, it is necessary to correctly set the \texttt{chain\_strength} parameter, which enforces all physical qubits within a chain to take the same value. 
If this value is too small, spin inconsistencies (chain breaks) are more likely to occur within the chain, reducing the reliability of the results. 
On the other hand, when implementing a QUBO on actual quantum annealing hardware, the coefficients assigned to spin variables are subject to an upper bound due to hardware limitations. 
If the \texttt{chain\_strength} is set excessively large, the original coefficients in the QUBO must be proportionally scaled down to fit within this range, which reduces the relative influence of the original objective terms and constraints. 
As a result, the problem representation on hardware may fail to reflect the intended optimization problem accurately. 
Therefore, selecting an appropriate \texttt{chain\_strength} is essential to preserve the balance between maintaining chain stability and faithfully encoding the original problem \cite{dwave-chainstrength}.
In this study, we experimented with different values of \texttt{chain\_strength} on several problem instances. We selected a balanced parameter value by comparing the stability of the obtained solutions and their optimization performance. 
When a chain break occurred, we applied D-Wave's \texttt{MinimizeEnergy} strategy, which automatically selects the most appropriate value from the perspective of energy minimization, thereby ensuring the consistency of the solutions.
Since QA is also a stochastic method, variability arises in the solutions obtained in each sampling. 
Therefore, as in SA, we set the number of samples to $\texttt{num\_reads} = 1000$ to suppress statistical fluctuations in the solutions. 
The parameter \texttt{annealing\_time}, which controls the execution time of quantum annealing, is analogous to \texttt{num\_sweeps} in SA. 
In this study, we fixed it at the default value of $\texttt{annealing\_time} = 20\,\mathrm{\mu s}$.

\section*{Results}
First, we compare the results of solving the naive formulation~\eqref{eq:qubo_objective_ideal}, which provides the exact optimal matching solution, using QA and SA, and evaluate the performance trends of each solver. 
In particular, for QA, we compare the solutions obtained with and without post-processing using the greedy method \texttt{SteepestDescentSolver} \cite{Greedy}, in order to examine the effect of post-processing on solution quality.
In this experiment, the numbers of users and supporters are set equal at $n$, and we consider a one-to-one matching problem. 
To clarify the scaling characteristics concerning the problem size $n$, no prior edge filtering is performed, and edges are added for all possible combinations of users and supporters.
In addition, the matching scores between user-supporter pairs were randomly assigned from a Gaussian distribution with a mean of 12.3 and a variance of 2.80, which is empirically estimated based on real-world data collected during the experiments conducted in Sendai City.
Consequently, the number of graph edges, and thus the number of required decision variables, is given by $|E|=|U\times S|=n^2$.
The upper limit of $n$ is set to 15, corresponding to a problem size close to the largest scale for which embedding is practically feasible on the \texttt{Advantage\_system6.4}.
\begin{figure}[htbp]
  \centering
  \includegraphics[width=0.7\textwidth]{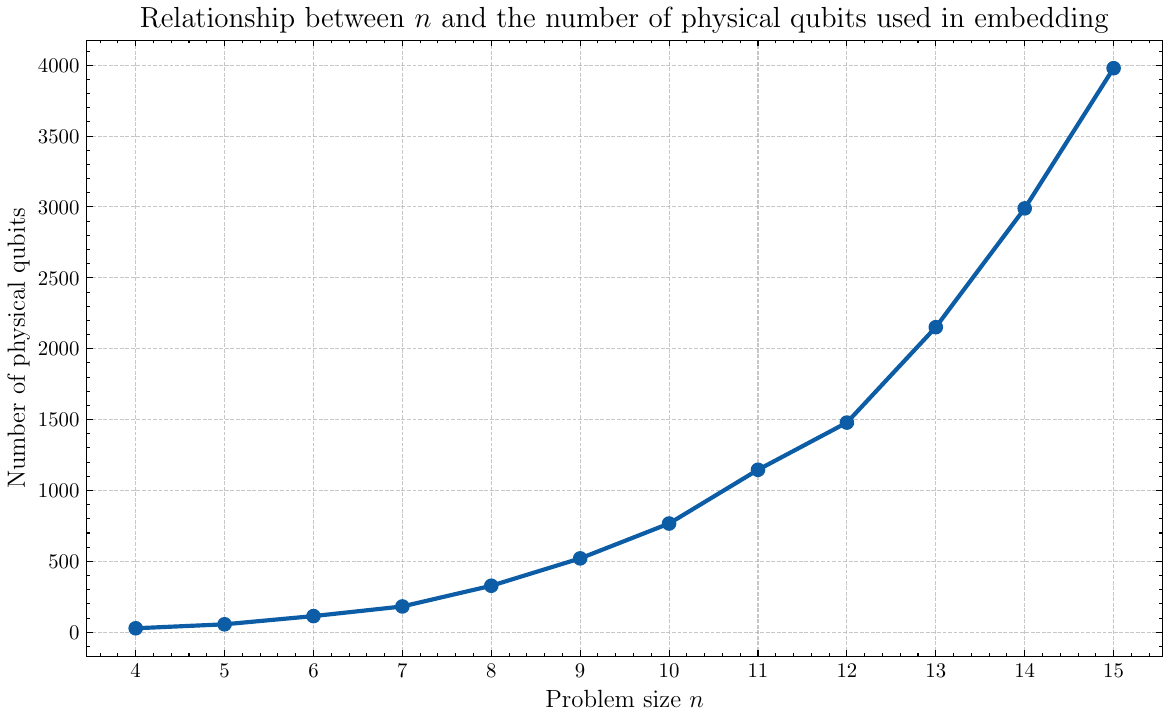}
  \caption{Comparison of the number of qubits required for embedding concerning problem size $n$.
  The horizontal axis shows the number of users and supporters $n$, and the vertical axis shows the number of qubits required for embedding.
  }
  \label{fig:n_vs_embedding}
\end{figure}
Figure~\ref{fig:n_vs_embedding} shows the change in the number of physical qubits required for embedding as a function of the problem size $n$.
Since the number of qubits available on the \texttt{Advantage\_system6.4} is approximately 5600, it can be seen that $n=15$ is close to the maximum size for which embedding is feasible.

\begin{figure}[htbp]
  \centering
  \includegraphics[width=0.7\textwidth]{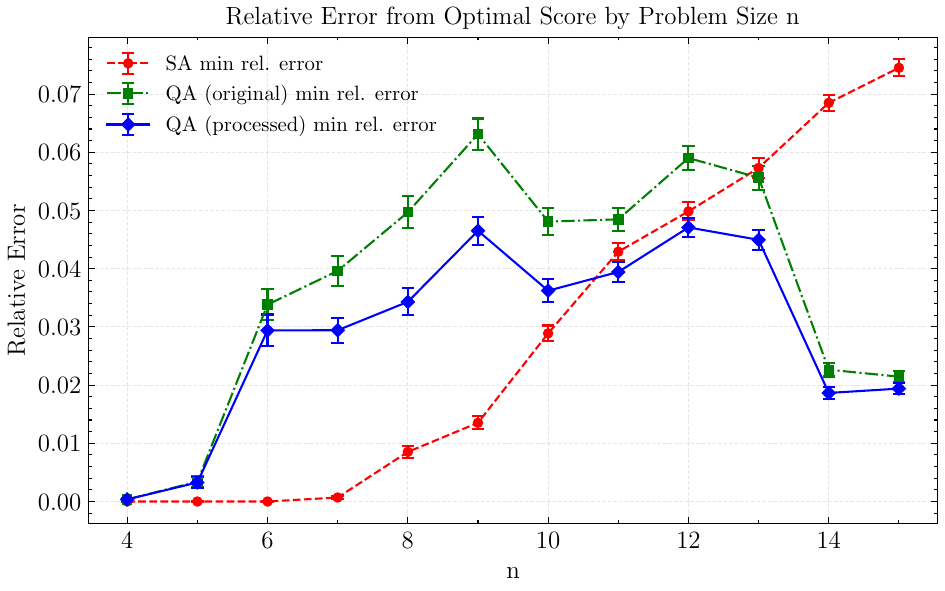}
  \caption{
    Comparison of the relative errors from the optimal solutions obtained using the exact solver Gurobi Optimizer (version 12.0.2).
    The horizontal axis represents the number of users and supporters $n$, and the vertical axis shows the relative error from the solutions obtained by each solver with respect to the optimal solution.
    Among the \texttt{num\_reads} samples, the relative error of the best feasible solution is shown in this figure.
    For each $n$, 100 random instances were generated, and the plotted values represent the mean and standard error across these instances.
  }
  \label{fig:n_vs_relative_score_naive}
\end{figure}
Figure~\ref{fig:n_vs_relative_score_naive} shows how the performance of different QUBO solvers changes in terms of finding optimal solutions.
The horizontal axis represents the number of users and supporters $n$, equal to the number of possible user–supporter pairs. 
The vertical axis indicates the relative error of the solutions obtained by each solver, calculated as $(E^{*}-E) / E^{*}$.
Here, $E$ denotes the cost computed from the solution $x=\{x_e\}_{e\in E}$ obtained by each method, using the objective function $\sum_{e\in E}M_ex_e$,
and $E^{*}$ is the optimal cost computed from the exact solution of Equation~\eqref{eq:qubo_objective_ideal}.

When using SA, the relative error of the best solutions tended to increase as the problem size $n$ increased.
For relatively small problem sizes, the optimal or near-optimal solutions could be sampled. 
However, for larger problem sizes, even the best solutions exhibited higher relative errors, showing a clear trend of diverging from the optimal solution.

In the case of QA, no trend of increasing error with problem size was observed, and especially for $n=14$ and $n=15$, tiny relative errors were obtained regardless of whether post-processing was applied.
These results indicate that while SA samples higher-quality solutions than QA for small values of $n$, QA obtains better solutions as $n$ becomes larger.
Focusing on the effect of post-processing in QA, we observed that post-processing improved relative error in the small problem size range, where the relative error was relatively large.
The reason why QA exhibited particularly strong performance at $n=14$ and $n=15$ remains unclear.
One possible explanation is that specific characteristics of the problem instances or the embedding on hardware happened to favor QA in these cases.
However, this behavior might just be a coincidence and not something intrinsic to the algorithm. 
Further investigation is required to clarify whether the observed trend reflects a systematic advantage or simply a coincidental occurrence.

\begin{figure}[htbp]
  \centering
  \begin{subfigure}{0.45\linewidth}
    \includegraphics[width=\linewidth]{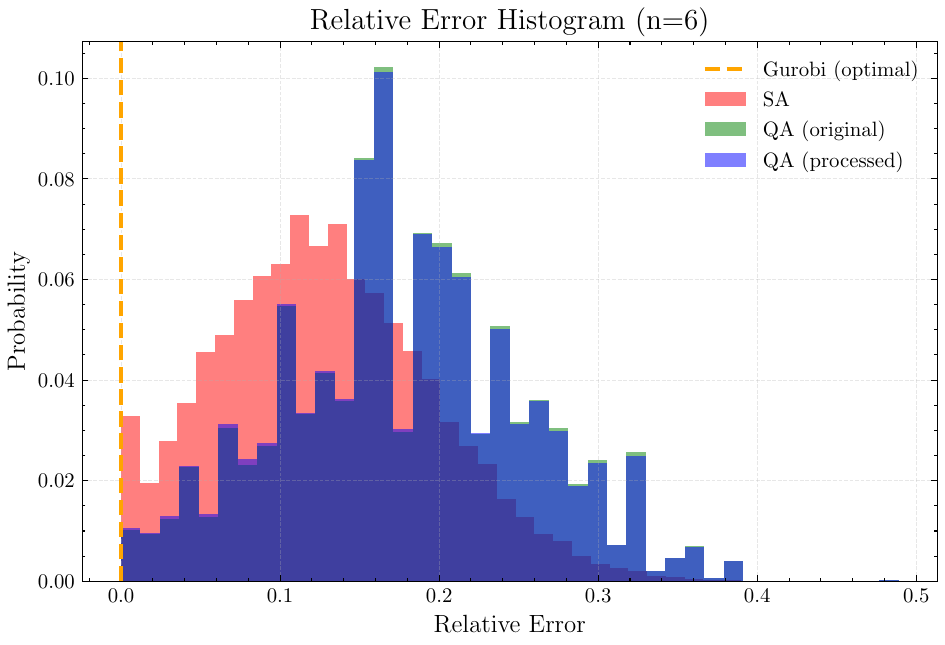}
    \caption{Energy histogram for $n=6$}
  \end{subfigure}
  \hspace{0.05\linewidth}
  \begin{subfigure}{0.45\linewidth}
    \includegraphics[width=\linewidth]{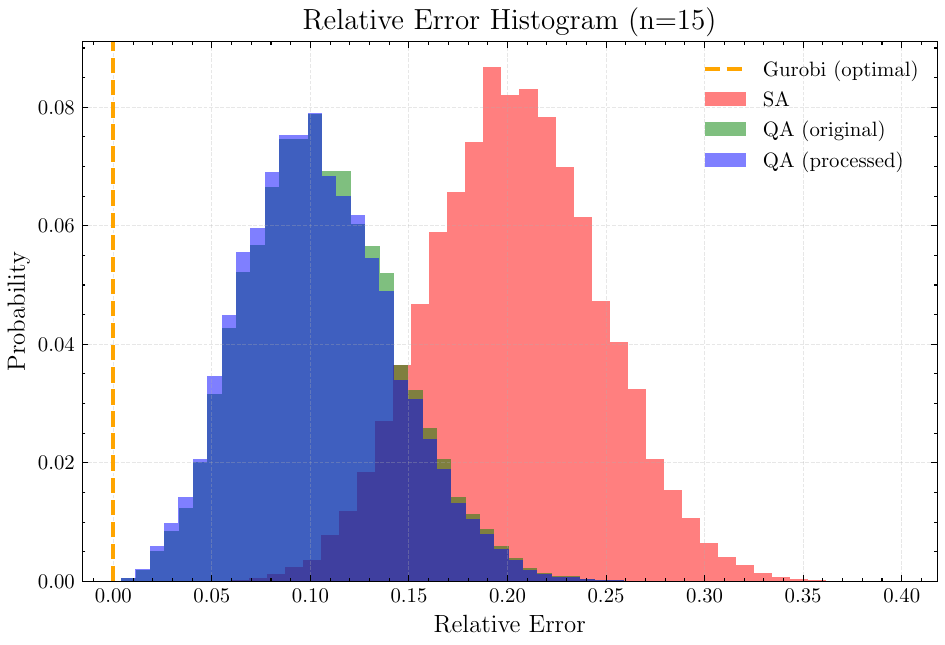}
    \caption{Energy histogram for $n=15$}
  \end{subfigure}
  \caption{
    Histograms of the energy distributions of the solutions obtained by SA and QA.
    The horizontal axis represents the relative error, and the vertical axis indicates the probability of occurrence of samples corresponding to each relative error.
    For QA, the results with and without post-processing are compared.
    The left-hand panel shows the results for $n=6$, and the right-hand panel shows those for $n=15$.
  }
  \label{fig:hist_relerror}
\end{figure}
Figure~\ref{fig:hist_relerror} shows histograms of the energy distributions of the samples obtained by SA and QA, with results presented for $n=6$ as a small-scale problem size and $n=15$ as a large-scale problem size.
For $n=6$, multiple high-quality samples with relative errors close to zero were observed in both SA and QA.
In particular, SA yielded a significantly larger samples matching the optimal solution, and its distribution was skewed toward the left.
In contrast, the distribution for QA was comparatively broader, including many samples with larger relative errors. 
Post-processing with the greedy method produced only a limited improvement in this problem size, and no significant effect was observed.
However, when the problem size was increased to $n=15$, the relative error distribution for SA shifted significantly to the right, clearly indicating a degradation in solution accuracy.
In contrast, the QA distribution shifted further to the left compared with the case of $n=6$, while becoming sharper, indicating reduced variability in the solutions.
While SA failed to obtain any solutions close to the optimal solution with relative errors below $0.05$, QA was able to sample such high-quality solutions.
Furthermore, the effect of post-processing became more pronounced for QA at this larger problem size, shifting the entire distribution further to the left and increasing the proportion of solutions with smaller relative errors. 
These results demonstrate that as the problem size increases, SA tends to experience performance degradation as the problem size increases, whereas QA benefits from improved solution stability and post-processing effects.

Next, we focus on solution diversity \cite{zucca2021diversitymetricevaluationquantum}, a metric that quantitatively evaluates how different multiple candidate solutions near the exact optimum are, and examine the tendencies exhibited by each solver. 
Finding the exact optimal solution is certainly important in the matching problem considered here. However, obtaining multiple high-quality solutions is of even greater value in practical applications.
This is because the solution with the highest optimization score is not necessarily the most suitable in practice in matching based on subjective and multidimensional criteria such as compatibility. 
Rather, it is desirable to have flexibility in decision-making by choosing from multiple viable options. 
In such situations, evaluating solvers using solution diversity is highly effective.
Diversity is determined as follows. First, from the feasible solutions generated by each solver, we extract the solutions that satisfy the allowable range
\begin{equation}
  E \leq E^{*} + \alpha(E_{\text{best}} - E^{*}),
\end{equation}
where $E_{\text{best}}$ is the best cost obtained by each solver, and $E^{*}$ is the optimal cost obtained by Gurobi.
Then, the difference between two solutions is defined by the Hamming distance, and two solutions are considered distinct if the distance exceeds $Rn$,
$n$ is the number of users and supporters, and $R$ is a threshold parameter determining the degree of distinctness.
Finally, diversity is defined as the size of the maximum independent set of such distinct solutions.

\begin{figure}[htbp]
  \centering
  \includegraphics[width=0.75\linewidth]{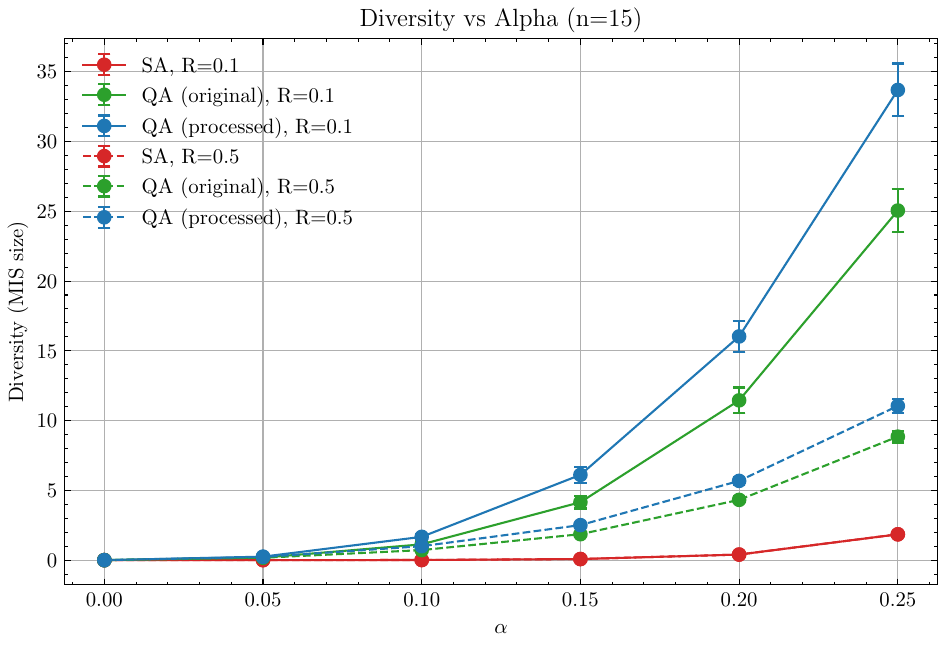}
    \caption{
    Each solver obtained a comparison of the solution diversity (size of the maximum independent set).  
    The horizontal axis \( \alpha \) represents the allowable error from the optimal cost, and the vertical axis indicates the diversity of the solution set within that range.  
    The solid and dashed lines show the differences for threshold parameter settings \( R = 0.1 \) and \( R = 0.5 \), respectively.
  }
  \label{fig:diversity}
\end{figure}

Figure~\ref{fig:diversity} shows the solution diversity obtained by SA and QA for $n=15$.
Even in the region where the allowable error $\alpha$ is significant, SA exhibits extremely low diversity, indicating that it is effectively unable to generate valid candidate solutions.
This stems from the fact that SA does not sufficiently sample solutions near the optimal solution, as observed in Figure~\ref{fig:hist_relerror}.
In contrast, QA can obtain a larger number of high-quality solutions, and because these solutions are also structurally different, its diversity scores are evaluated to be higher.
In particular, in the region $\alpha \geq 0.15$, the solutions obtained with post-processing demonstrate significantly greater diversity than those obtained by the other methods, confirming their effectiveness by providing flexible options in practical applications.

\begin{figure}[htbp]
  \centering
  \includegraphics[width=0.75\linewidth]{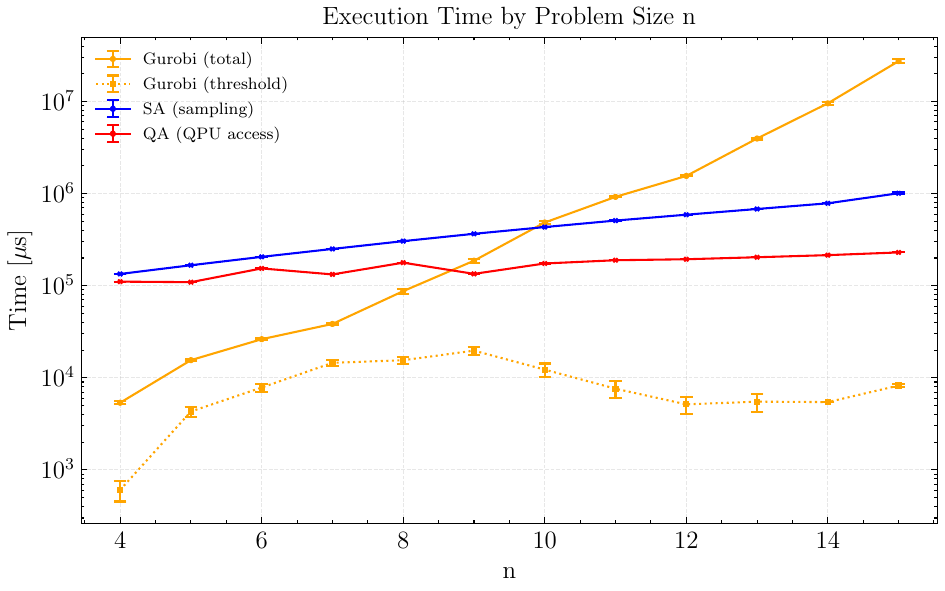}
  \caption{
    Comparison of computation times for each solver. The horizontal axis represents the problem size \( n \), and the vertical axis shows the computation time.  
    QA (QPU access) and SA (sampling) represent the time required to execute the predefined 1000 samples for a single problem instance and obtain the results.  
    Gurobi (total) represents the total time required to find the optimal solution to the QUBO, including the time needed to prove solution optimality.  
    Gurobi (threshold) indicates the time required for Gurobi to obtain a solution with the same cost as the best solution found by SA for the same instance.
  }
  \label{fig:time_vs_n}
\end{figure}

Figure~\ref{fig:time_vs_n} compares the computation times required to solve the QUBO using QA, SA, and Gurobi as a function of the problem size. 
The QPU access time for QA is the sum of the QPU sampling time and the programming time. The sampling time includes the annealing time, readout time, and hardware delays required for all samples. 
In contrast, the programming time occurs only once per problem instance and corresponds to configuring the QUBO on the quantum chip.
In this study, we treat the QPU access time, which represents the latency observed in real-world applications, as the computation time of QA.

The total computation time of Gurobi increases significantly with problem size, requiring approximately $20\,\mathrm{s}$ to solve a single instance when $n=15$.
However, the time required for Gurobi to find the best solution obtained by SA is extremely short, remaining around $10\,\mathrm{ms}$ for all problem sizes.
This indicates that most of Gurobi's total computation time is spent on proving the optimality of the obtained solution.
Because QA and SA do not perform optimality proofs, comparing their computation times with Gurobi (threshold) is more appropriate. 
From this perspective, SA is the slowest among the three methods, with computation time increasing exponentially with problem size and requiring approximately $1\,\mathrm{s}$ when $n=15$.
QA lies between the two; the total annealing time for 1000 samples is only about $20\,\mathrm{ms}$, but when overheads such as initialization and readout for QPU access are included,
the overall computation time amounts to approximately $0.1\,\mathrm{s}$. The increase in computation time with problem size is also more moderate for QA than for SA.
Among the three methods, Gurobi can search for solutions the fastest, but when latency due to proving optimality is also considered, it results in the longest computation time.

\begin{figure}[htbp]
  \centering
  \includegraphics[width=0.75\textwidth]{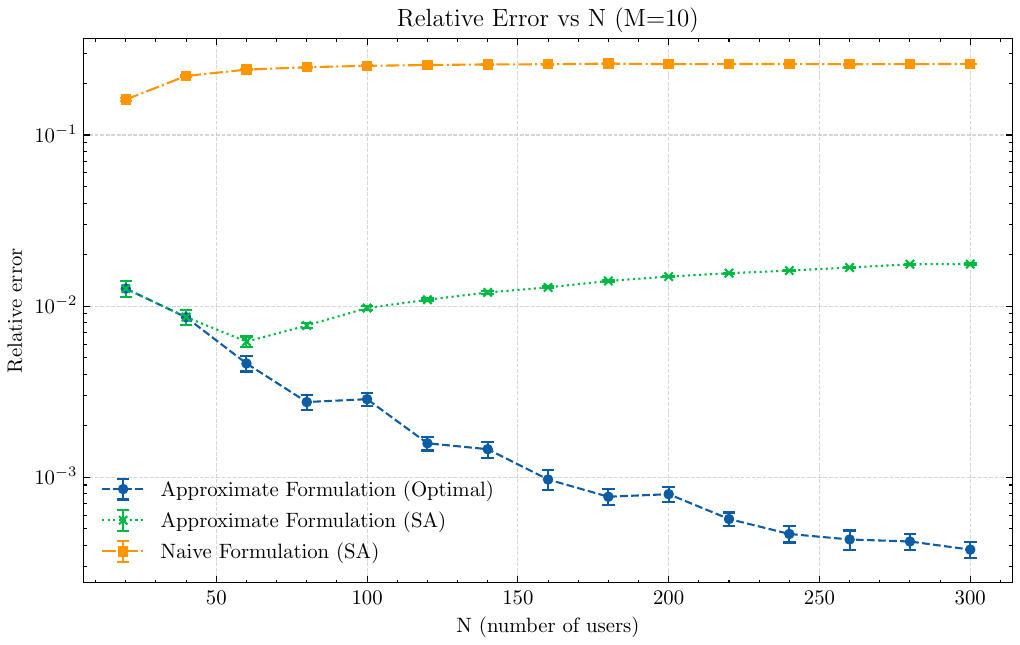}
  \caption{
    The relative error of the two formulations in SA depends on the number of users, with the number of supporters fixed at $M=10$.
    Here, the relative error is defined as the deviation from the optimal solution of the naive formulation~\eqref{eq:qubo_objective_ideal}.
  }
  \label{fig:relerr_vs_N_10}
\end{figure}

\begin{figure}[htbp]
  \centering
  \includegraphics[width=0.75\textwidth]{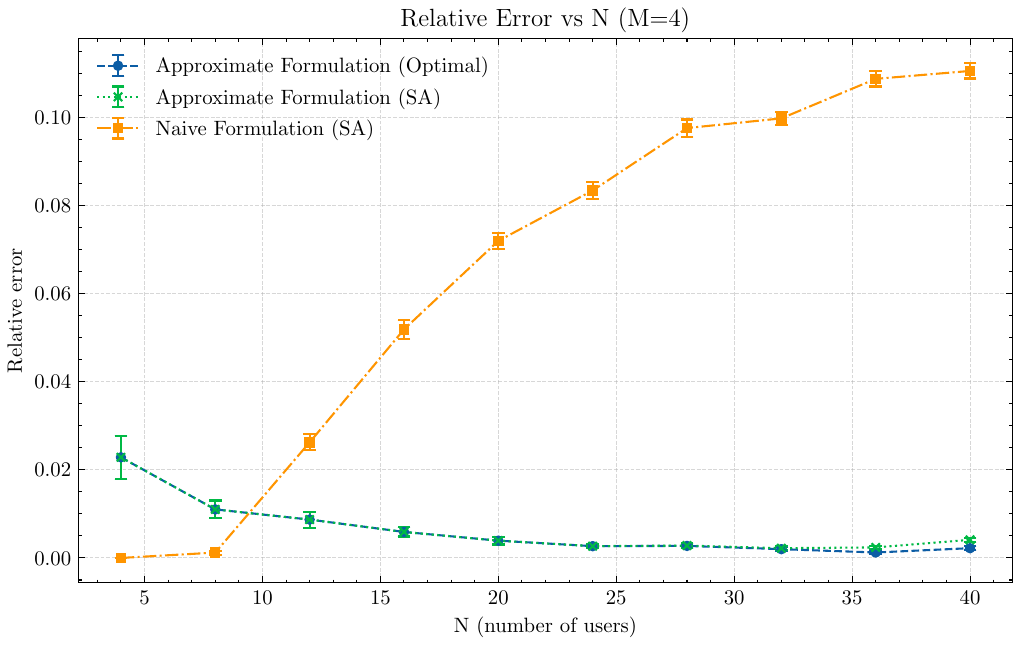}
  \caption{
    The relative error of the two formulations in SA depends on the number of users, with the number of supporters fixed at $M=4$.
    Here, the relative error is defined as the deviation from the optimal solution of the naive formulation~\eqref{eq:qubo_objective_ideal}.
  }
  \label{fig:relerr_vs_N_4}
\end{figure}

Next, we discuss the accuracy of solutions obtained in SA when using the approximate formulation~\eqref{eq:qubo_objective_approx}, which is expected to provide good results for large-scale problems. 
In this experiment, we varied the number of users $N$ while fixing the number of supporters $M$. 
We compared the relative errors of the solutions obtained using the naive formulation~\eqref{eq:qubo_objective_ideal} and the approximate formulation. 
Since the approximate formulation is constructed by further reducing edges from the naive formulation, the optimal solution score obtained by the approximate formulation is always less than or equal to that of the naive formulation.
At the same time, the approximate formulation requires fewer decision variables when representing the same problem instance, which implies that, under heuristic algorithms such as SA, the solutions obtainable within a practical computational time may yield better scores with the approximate formulation than with the naive formulation.
This tendency is demonstrated by the experimental results shown in Figure~\ref{fig:relerr_vs_N_10} and Figure~\ref{fig:relerr_vs_N_4}.

The results for the larger-scale setting with $M=10$ are shown in Figure~\ref{fig:relerr_vs_N_10}.
The approximation bound of the formulation is indicated, and the results demonstrate that its approximation accuracy improves as the number of users $N$ increases.
Furthermore, when comparing the solutions obtained by SA for each formulation, the scores achieved with the approximate formulation consistently exhibit better accuracy than those obtained with the naive formulation.
The results for the smaller-scale setting with $M=4$ are presented in Figure~\ref{fig:relerr_vs_N_4}.
For $N=5$ and $N=10$, where SA can efficiently explore solutions near the optimum of the naive formulation, the limitation of the approximation accuracy inherent in the approximate formulation becomes a bottleneck, and the naive formulation yields better results.
However, as $N$ increases, the solution accuracy of the naive formulation becomes worse, whereas the approximate formulation continues to produce nearly optimal solutions.
This behavior can be understood from the property that the number of possible pairings becomes sufficiently large relative to the required number of assignments.
When compatibility scores are determined at random, the probability that a given user $i$ selects a supporter $j$ as one of their candidate list is $2/M$.
Therefore, the expected number of users who include supporter $j$ in their candidate list is $2N/M$.
In contrast, the number of users that must actually be assigned to supporter $j$ is only $N/M$ if the assignments are evenly distributed across supporters.
Thus, it suffices to select $N/M$ highly compatible users from the pool of $2N/M$ candidates.
Consequently, as $N/M$ increases, the surplus of candidates also increases, making the optimal solution under the naive formulation more likely to coincide with the solution obtained using the approximate formulation.
In cases such as $N=M$, however, the approximate formulation provides only about two candidates in expectation, forcing a binary choice and thereby increasing the likelihood of discrepancies from the optimum of the naive formulation.
This indicates that, while the limitation of the approximation is more evident for very small problem instances, the approximate formulation remains stable in exploring near-optimal solutions as the problem grows, yielding better scores than the naive formulation.

Finally, we demonstrate the results of applying the proposed formulation to an actual matching problem. 
Participants were recruited in Sendai City, and a matching was conducted between parents raising children in the city and senior residents. 
14 users and 14 supporters participated as matching candidates, and a one-to-one matching was performed.
\begin{figure}[htbp]
  \centering
  \includegraphics[width=0.5\textwidth]{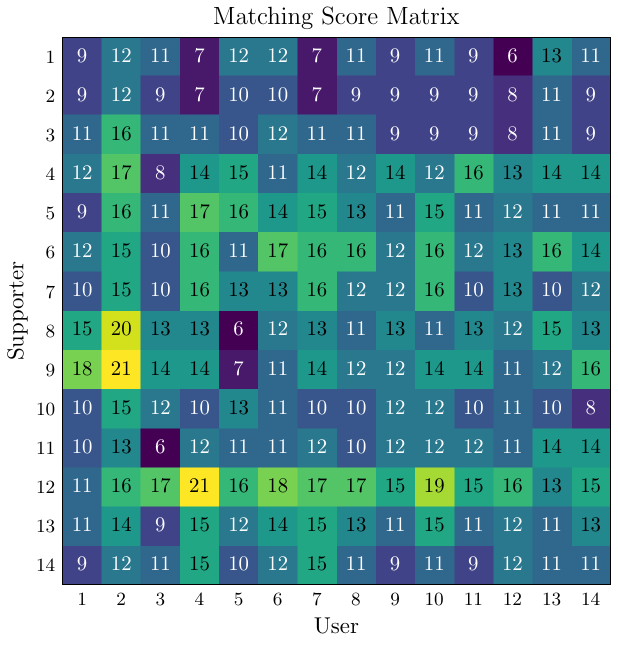}
  \caption{
    Compatibility scores between pairs are calculated based on the actual questionnaire results.  
    There are 14 users and 14 supporters, and a compatibility score is assigned to each pair.  
    The questionnaire consists of 10 items, and since the maximum item score \( m_{ij}^{\mu} \) is three, the maximum compatibility score \( M_{ij} \) is 30.
    In the present questionnaire results, the maximum score was 21 and the minimum score was six.
  }
  \label{fig:sendai_matching_matrix}
\end{figure}
\begin{figure}[htbp]
  \centering
  \includegraphics[width=0.7\textwidth]{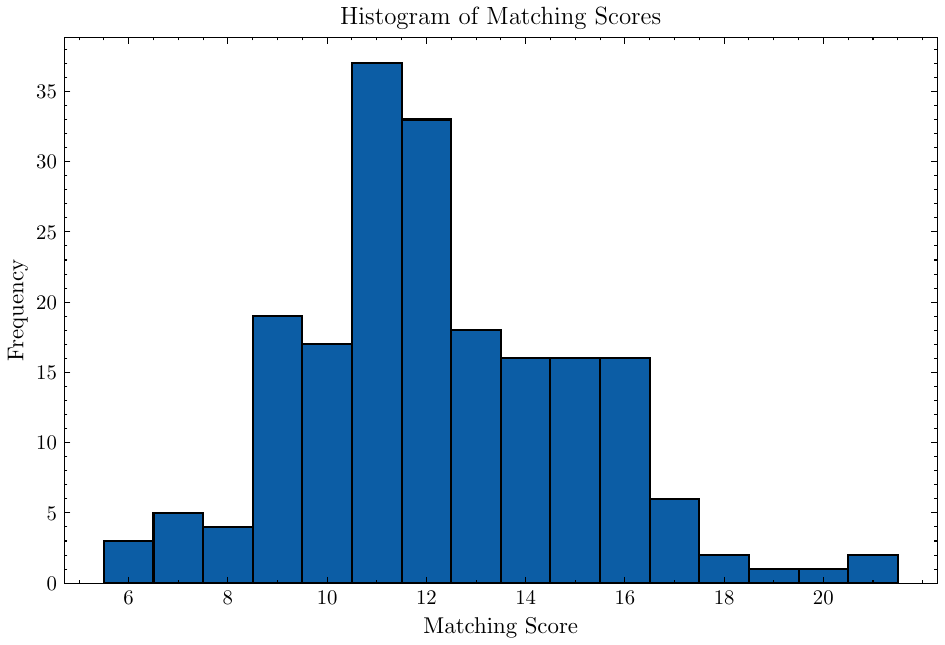}
  \caption{
    Histogram visualizing the distribution of compatibility scores.  
    There are few extremely compatible or incompatible pairs, and the distribution shows a gradual peak around 11 and 12.
  }
  \label{fig:sendai_matching_histogram}
\end{figure}
The compatibility scores for each pair, calculated based on the actual questionnaire responses, are shown in Figure~\ref{fig:sendai_matching_matrix}, and the distribution is visualized as a histogram in Figure~\ref{fig:sendai_matching_histogram}. 
When the Gaussian distribution approximates the histogram, the estimated mean and variance are 12.3 and 2.80, respectively. 
As described previously, the matching scores used in the experiments with random instances follow a Gaussian distribution with these mean and variance values.
\begin{figure}[htbp]
  \centering
  \includegraphics[width=0.5\textwidth]{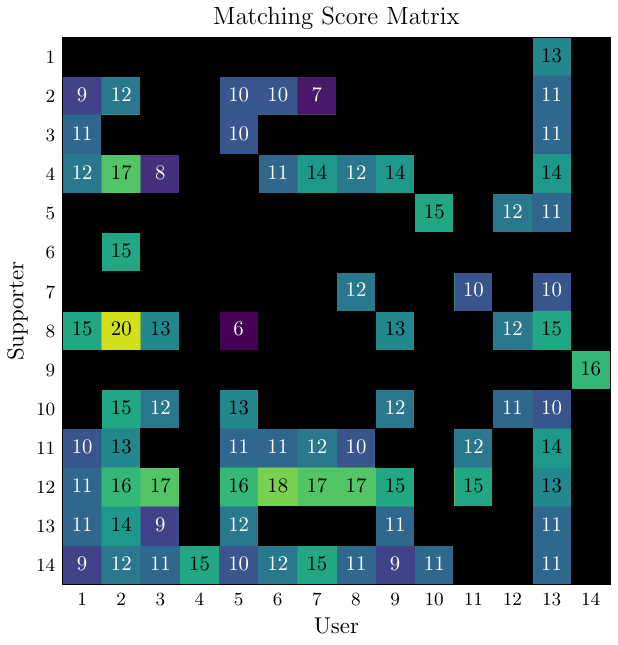}
  \caption{
    The compatibility score matrix is after masking the infeasible pairs for matching due to the pre-processing filter.  
    Approximately \SI{38}{\percent} of the pairs remain feasible for matching. However, a one-to-one matching can still be achieved.
  }
  \label{fig:sendai_matching_matrix_masked}
\end{figure}

\begin{figure}[htbp]
  \centering
  \includegraphics[width=0.5\textwidth]{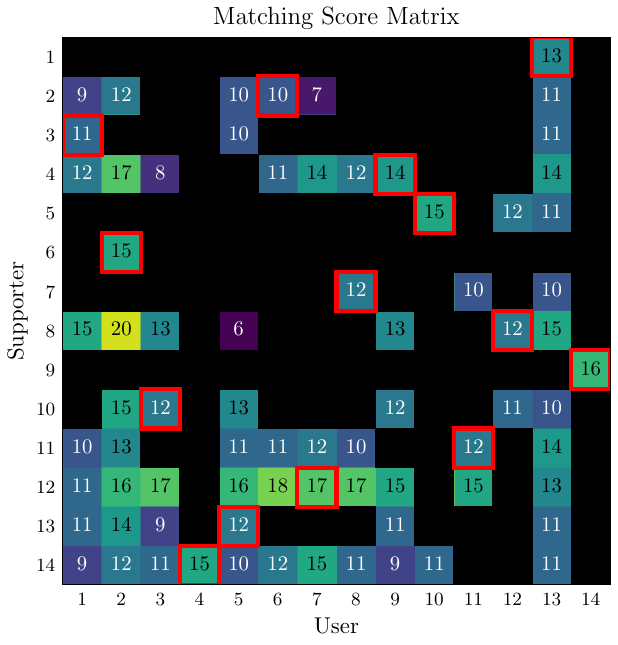}
  \caption{
    One of the optimal matchings obtained from sampling with SA.  
    The pairs enclosed in red boxes represent the matched pairs, with exactly one user assigned to each supporter.  
    Among the 1000 samples, all eight optimal matching solutions were successfully obtained.
  }
  \label{fig:res1}
\end{figure}
Figure~\ref{fig:sendai_matching_matrix_masked} shows the compatibility score matrix from Figure~\ref{fig:sendai_matching_matrix} with the pairs excluded by filtering masked out, and even after filtering, it was still possible to achieve a perfect matching despite approximately 60\% of the pairs being excluded.
Each supporter was strictly required to be assigned exactly one user in this demonstration experiment. 
Therefore, because the approximate formulation~\eqref{eq:qubo_objective_approx}, which allows for more efficient use of decision variables, does not necessarily satisfy this constraint and because filtering reduced the total number of pairs to 74, making the problem size sufficiently small, we used the exact formulation~\eqref{eq:qubo_objective_ideal} and performed optimization using SA.

Under the current filtering conditions, there were eight distinct matchings with the optimal score, and the optimization using SA could sample all eight of these matchings.
One of the matching results is shown in Figure~\ref{fig:res1}.
None of the matched pairs included extremely incompatible pairs with compatibility scores below 10, and most pairs had compatibility scores higher than the mode of the matching scores, which was 11.
It was also confirmed that one user was assigned to each supporter, achieving a one-to-one matching.
The ability to obtain multiple optimal matchings implies several options for creating visit schedules, suggesting that flexible responses can be made in actual operations.

\section*{Conclusion}
In this study, we demonstrated a compatibility-based matching framework that formulates the assignment of parents to experienced senior supporters as a QUBO model and benchmarked QA against SA.
Across randomized instances scaled to realistic problem sizes, QA maintained solution quality as problem size increased and, crucially, sampled a broader set of near-optimal matchings than SA.
While SA performed strongly on small instances, its quality and diversity degraded with scale.
These results indicate a robust sampling advantage for QA in operations where flexibility among several high-quality alternatives is more valuable than a single exact optimum.

We also introduced a compressed "top-2" approximate formulation~\eqref{eq:qubo_objective_approx} that reduces the number of decision variables to the number of users while preserving high matching scores.
By using this formulation, we can reduce the number of decision variables required to solve the matching problem while still obtaining high-quality solutions, thereby enabling us to handle larger matching problems that are difficult to solve with the naive formulation.
Furthermore, this approximation eased parameter tuning and computational requirements and produced solutions close to the exact formulation, making it easier to apply in practice when resources are limited.
Beyond randomized and large-scale simulations, we conducted a field study in Sendai City to evaluate the effectiveness of the proposed framework in a real-world setting.
A field study in Sendai City demonstrated feasibility in practice; after pre-filtering infeasible pairs, the system generated multiple optimal one-to-one matchings, and many distinct optima were obtained.
The existence of multiple optimal matchings provided flexibility in arranging visit schedules. As a result, alternative assignments could be selected to accommodate practical constraints without compromising compatibility.

Future work will extend the model by using QUBO interaction terms, which allow the formulation to capture compatibility effects that only appear when multiple assignments occur together.
For instance, parents with similar personalities or concerns could be encouraged to be matched with the same supporter, reducing mismatches and enabling more consistent support.
Such extensions would make it possible to incorporate more complex matching criteria into the formulation, going beyond one-to-one compatibility to account for group-level interactions.
In doing so, the optimization framework could express richer objectives.

\bibliography{sendai}

\begin{thebibliography}{10}
\urlstyle{rm}
\expandafter\ifx\csname url\endcsname\relax
  \def\url#1{\texttt{#1}}\fi
\expandafter\ifx\csname urlprefix\endcsname\relax\def\urlprefix{URL }\fi
\expandafter\ifx\csname doiprefix\endcsname\relax\def\doiprefix{DOI: }\fi
\providecommand{\bibinfo}[2]{#2}
\providecommand{\eprint}[2][]{\url{#2}}

\bibitem{naikaku}
\bibinfo{author}{{Cabinet Secretariat, Office for the Establishment of the Children and Families Agency}}.
\newblock \bibinfo{title}{Current situation of children and childrearing and the voices and attitudes of youth and childrearing stakeholders}.
\newblock \bibinfo{howpublished}{\url{https://www.cas.go.jp/jp/seisaku/kodomo_seisaku_kyouka/dai1/siryou5.pdf}} (\bibinfo{year}{2023}).
\newblock \bibinfo{note}{Accessed: July 23, 2025}.

\bibitem{morita2021childcare}
\bibinfo{author}{Morita, M.}, \bibinfo{author}{Saito, A.}, \bibinfo{author}{Nozaki, M.} \& \bibinfo{author}{Ihara, Y.}
\newblock \bibinfo{journal}{\bibinfo{title}{Childcare support and child social development in japan: investigating the mediating role of parental psychological condition and parenting style}}.
\newblock {\emph{\JournalTitle{Philosophical Transactions of the Royal Society B: Biological Sciences}}} \textbf{\bibinfo{volume}{376}}, \bibinfo{pages}{20200025}, \doiprefix\url{10.1098/rstb.2020.0025} (\bibinfo{year}{2021}).

\bibitem{Glover2022}
\bibinfo{author}{Glover, F.}, \bibinfo{author}{Kochenberger, G.}, \bibinfo{author}{Hennig, R.} \& \bibinfo{author}{Du, Y.}
\newblock \bibinfo{journal}{\bibinfo{title}{Quantum bridge analytics i: a tutorial on formulating and using qubo models}}.
\newblock {\emph{\JournalTitle{Annals of Operations Research}}} \textbf{\bibinfo{volume}{314}}, \bibinfo{pages}{141--183}, \doiprefix\url{10.1007/s10479-022-04634-2} (\bibinfo{year}{2022}).

\bibitem{kadowaki1998}
\bibinfo{author}{Kadowaki, T.} \& \bibinfo{author}{Nishimori, H.}
\newblock \bibinfo{journal}{\bibinfo{title}{Quantum annealing in the transverse ising model}}.
\newblock {\emph{\JournalTitle{Phys. Rev. E}}} \textbf{\bibinfo{volume}{58}}, \bibinfo{pages}{5355--5363}, \doiprefix\url{10.1103/PhysRevE.58.5355} (\bibinfo{year}{1998}).

\bibitem{lucas2014ising}
\bibinfo{author}{Lucas, A.}
\newblock \bibinfo{journal}{\bibinfo{title}{Ising formulations of many np problems}}.
\newblock {\emph{\JournalTitle{Frontiers in Physics}}} \textbf{\bibinfo{volume}{Volume 2 - 2014}}, \doiprefix\url{10.3389/fphy.2014.00005} (\bibinfo{year}{2014}).

\bibitem{neukart2017trafficflowoptimizationusing}
\bibinfo{author}{Neukart, F.} \emph{et~al.}
\newblock \bibinfo{title}{Traffic flow optimization using a quantum annealer} (\bibinfo{year}{2017}).
\newblock \eprint{1708.01625}.

\bibitem{Inoue2021}
\bibinfo{author}{Inoue, D.}, \bibinfo{author}{Okada, A.}, \bibinfo{author}{Matsumori, T.}, \bibinfo{author}{Aihara, K.} \& \bibinfo{author}{Yoshida, H.}
\newblock \bibinfo{journal}{\bibinfo{title}{Traffic signal optimization on a square lattice with quantum annealing}}.
\newblock {\emph{\JournalTitle{Scientific Reports}}} \textbf{\bibinfo{volume}{11}}, \bibinfo{pages}{3303}, \doiprefix\url{10.1038/s41598-021-82740-0} (\bibinfo{year}{2021}).

\bibitem{shikanai2023}
\bibinfo{author}{Shikanai, R.}, \bibinfo{author}{Ohzeki, M.} \& \bibinfo{author}{Tanaka, K.}
\newblock \bibinfo{journal}{\bibinfo{title}{Quadratic unconstrained binary formulation for traffic signal optimization on real-world maps}}.
\newblock {\emph{\JournalTitle{Journal of the Physical Society of Japan}}} \textbf{\bibinfo{volume}{94}}, \bibinfo{pages}{024001}, \doiprefix\url{10.7566/JPSJ.94.024001} (\bibinfo{year}{2025}).

\bibitem{ohzeki2019control}
\bibinfo{author}{Ohzeki, M.}, \bibinfo{author}{Miki, A.}, \bibinfo{author}{Miyama, M.~J.} \& \bibinfo{author}{Terabe, M.}
\newblock \bibinfo{journal}{\bibinfo{title}{Control of automated guided vehicles without collision by quantum annealer and digital devices}}.
\newblock {\emph{\JournalTitle{Frontiers in Computer Science}}} \textbf{\bibinfo{volume}{Volume 1 - 2019}}, \doiprefix\url{10.3389/fcomp.2019.00009} (\bibinfo{year}{2019}).

\bibitem{Haba2022}
\bibinfo{author}{Haba, R.}, \bibinfo{author}{Ohzeki, M.} \& \bibinfo{author}{Tanaka, K.}
\newblock \bibinfo{journal}{\bibinfo{title}{Travel time optimization on multi-agv routing by reverse annealing}}.
\newblock {\emph{\JournalTitle{Scientific Reports}}} \textbf{\bibinfo{volume}{12}}, \bibinfo{pages}{17753}, \doiprefix\url{10.1038/s41598-022-22704-0} (\bibinfo{year}{2022}).

\bibitem{rosenberg2016solving}
\bibinfo{author}{Rosenberg, G.} \emph{et~al.}
\newblock \bibinfo{journal}{\bibinfo{title}{Solving the optimal trading trajectory problem using a quantum annealer}}.
\newblock {\emph{\JournalTitle{IEEE Journal of Selected Topics in Signal Processing}}} \textbf{\bibinfo{volume}{10}}, \bibinfo{pages}{1053--1060}, \doiprefix\url{10.1109/JSTSP.2016.2574703} (\bibinfo{year}{2016}).

\bibitem{Venturelli2019}
\bibinfo{author}{Venturelli, D.} \& \bibinfo{author}{Kondratyev, A.}
\newblock \bibinfo{journal}{\bibinfo{title}{Reverse quantum annealing approach to portfolio optimization problems}}.
\newblock {\emph{\JournalTitle{Quantum Machine Intelligence}}} \textbf{\bibinfo{volume}{1}}, \bibinfo{pages}{17--30}, \doiprefix\url{10.1007/s42484-019-00001-w} (\bibinfo{year}{2019}).

\bibitem{yonaga2022quantum}
\bibinfo{author}{Yonaga, K.} \emph{et~al.}
\newblock \bibinfo{journal}{\bibinfo{title}{Quantum optimization with lagrangian decomposition for multiple-process scheduling in steel manufacturing}}.
\newblock {\emph{\JournalTitle{ISIJ International}}} \textbf{\bibinfo{volume}{62}}, \bibinfo{pages}{1874--1880}, \doiprefix\url{10.2355/isijinternational.ISIJINT-2022-019} (\bibinfo{year}{2022}).

\bibitem{PhysRevX.8.021050}
\bibinfo{author}{Amin, M.~H.}, \bibinfo{author}{Andriyash, E.}, \bibinfo{author}{Rolfe, J.}, \bibinfo{author}{Kulchytskyy, B.} \& \bibinfo{author}{Melko, R.}
\newblock \bibinfo{journal}{\bibinfo{title}{Quantum boltzmann machine}}.
\newblock {\emph{\JournalTitle{Phys. Rev. X}}} \textbf{\bibinfo{volume}{8}}, \bibinfo{pages}{021050}, \doiprefix\url{10.1103/PhysRevX.8.021050} (\bibinfo{year}{2018}).

\bibitem{OMalley2018}
\bibinfo{author}{O'Malley, D.}, \bibinfo{author}{Vesselinov, V.~V.}, \bibinfo{author}{Alexandrov, B.~S.} \& \bibinfo{author}{Alexandrov, L.~B.}
\newblock \bibinfo{journal}{\bibinfo{title}{Nonnegative/binary matrix factorization with a d-wave quantum annealer}}.
\newblock {\emph{\JournalTitle{PLoS ONE}}} \textbf{\bibinfo{volume}{13}}, \bibinfo{pages}{e0206653}, \doiprefix\url{10.1371/journal.pone.0206653} (\bibinfo{year}{2018}).

\bibitem{Sato2021}
\bibinfo{author}{Sato, T.}, \bibinfo{author}{Ohzeki, M.} \& \bibinfo{author}{Tanaka, K.}
\newblock \bibinfo{journal}{\bibinfo{title}{Assessment of image generation by quantum annealer}}.
\newblock {\emph{\JournalTitle{Scientific Reports}}} \textbf{\bibinfo{volume}{11}}, \bibinfo{pages}{13523}, \doiprefix\url{10.1038/s41598-021-92295-9} (\bibinfo{year}{2021}).

\bibitem{Urushibata_2022}
\bibinfo{author}{Urushibata, M.}, \bibinfo{author}{Ohzeki, M.} \& \bibinfo{author}{Tanaka, K.}
\newblock \bibinfo{journal}{\bibinfo{title}{Comparing the effects of boltzmann machines as associative memory in generative adversarial networks between classical and quantum samplings}}.
\newblock {\emph{\JournalTitle{Journal of the Physical Society of Japan}}} \textbf{\bibinfo{volume}{91}}, \doiprefix\url{10.7566/jpsj.91.074008} (\bibinfo{year}{2022}).

\bibitem{hasegawa2023kernellearningquantumannealer}
\bibinfo{author}{Hasegawa, Y.}, \bibinfo{author}{Oshiyama, H.} \& \bibinfo{author}{Ohzeki, M.}
\newblock \bibinfo{title}{Kernel learning by quantum annealer} (\bibinfo{year}{2023}).
\newblock \eprint{2304.10144}.

\bibitem{Goto_2025}
\bibinfo{author}{Goto, T.} \& \bibinfo{author}{Ohzeki, M.}
\newblock \bibinfo{journal}{\bibinfo{title}{Online calibration scheme for training restricted boltzmann machines with quantum annealing}}.
\newblock {\emph{\JournalTitle{Journal of the Physical Society of Japan}}} \textbf{\bibinfo{volume}{94}}, \doiprefix\url{10.7566/jpsj.94.034002} (\bibinfo{year}{2025}).

\bibitem{doi:10.7566/JPSJ.94.054003}
\bibinfo{author}{Takabayashi, T.}, \bibinfo{author}{Goto, T.} \& \bibinfo{author}{Ohzeki, M.}
\newblock \bibinfo{journal}{\bibinfo{title}{Subgradient method using quantum annealing for inequality-constrained binary optimization problems}}.
\newblock {\emph{\JournalTitle{Journal of the Physical Society of Japan}}} \textbf{\bibinfo{volume}{94}}, \bibinfo{pages}{054003}, \doiprefix\url{10.7566/JPSJ.94.054003} (\bibinfo{year}{2025}).
\newblock \eprint{https://doi.org/10.7566/JPSJ.94.054003}.

\bibitem{yonaga2020solvinginequalityconstrainedbinaryoptimization}
\bibinfo{author}{Yonaga, K.}, \bibinfo{author}{Miyama, M.~J.} \& \bibinfo{author}{Ohzeki, M.}
\newblock \bibinfo{title}{Solving inequality-constrained binary optimization problems on quantum annealer} (\bibinfo{year}{2020}).
\newblock \eprint{2012.06119}.

\bibitem{KoukiYonaga2022ISIJINT-2022-019}
\bibinfo{author}{Yonaga, K.} \emph{et~al.}
\newblock \bibinfo{journal}{\bibinfo{title}{Quantum optimization with lagrangian decomposition for multiple-process scheduling in steel manufacturing}}.
\newblock {\emph{\JournalTitle{ISIJ International}}} \textbf{\bibinfo{volume}{62}}, \bibinfo{pages}{1874--1880}, \doiprefix\url{10.2355/isijinternational.ISIJINT-2022-019} (\bibinfo{year}{2022}).

\bibitem{Ohzeki2020}
\bibinfo{author}{Ohzeki, M.}
\newblock \bibinfo{journal}{\bibinfo{title}{Breaking limitation of quantum annealer in solving optimization problems under constraints}}.
\newblock {\emph{\JournalTitle{Scientific Reports}}} \textbf{\bibinfo{volume}{10}}, \bibinfo{pages}{3126}, \doiprefix\url{10.1038/s41598-020-60022-5} (\bibinfo{year}{2020}).

\bibitem{doi:10.7566/JPSJ.92.113002}
\bibinfo{author}{Hirama, S.} \& \bibinfo{author}{Ohzeki, M.}
\newblock \bibinfo{journal}{\bibinfo{title}{Efficient algorithm for binary quadratic problem by column generation and quantum annealing}}.
\newblock {\emph{\JournalTitle{Journal of the Physical Society of Japan}}} \textbf{\bibinfo{volume}{92}}, \bibinfo{pages}{113002}, \doiprefix\url{10.7566/JPSJ.92.113002} (\bibinfo{year}{2023}).
\newblock \eprint{https://doi.org/10.7566/JPSJ.92.113002}.

\bibitem{ohzeki2023qubits}
\bibinfo{author}{Ohzeki, M.}
\newblock \bibinfo{title}{Presentation at qubits 2023} (\bibinfo{year}{2023}).
\newblock \bibinfo{note}{In presentation at Qubits 2023 Conference}.

\bibitem{takabayashi2025optimization}
\bibinfo{author}{Takabayashi, T.}, \bibinfo{author}{Sudo, S.}, \bibinfo{author}{Aoki, T.}, \bibinfo{author}{Seo, S.} \& \bibinfo{author}{Ohzeki, M.}
\newblock \bibinfo{journal}{\bibinfo{title}{Optimization of connection patterns between mobile phones and base stations using quantum annealing}}.
\newblock {\emph{\JournalTitle{Scientific Reports}}} \textbf{\bibinfo{volume}{15}}, \doiprefix\url{10.1038/s41598-025-09230-5} (\bibinfo{year}{2025}).

\bibitem{SA}
\bibinfo{author}{Kirkpatrick, S.}, \bibinfo{author}{Gelatt, C.~D.} \& \bibinfo{author}{Vecchi, M.~P.}
\newblock \bibinfo{journal}{\bibinfo{title}{Optimization by simulated annealing}}.
\newblock {\emph{\JournalTitle{Science}}} \textbf{\bibinfo{volume}{220}}, \bibinfo{pages}{671--680}, \doiprefix\url{10.1126/science.220.4598.671} (\bibinfo{year}{1983}).
\newblock \eprint{https://www.science.org/doi/pdf/10.1126/science.220.4598.671}.

\bibitem{SASampler}
\bibinfo{author}{OpenJij}.
\newblock \bibinfo{title}{Openjij: Framework for the ising model and qubo}.
\newblock \bibinfo{howpublished}{\url{https://github.com/Jij-Inc/OpenJij}} (\bibinfo{year}{2023}).
\newblock \bibinfo{note}{Accessed: 2025-07-22}.

\bibitem{cai2014practicalheuristicfindinggraph}
\bibinfo{author}{Cai, J.}, \bibinfo{author}{Macready, W.~G.} \& \bibinfo{author}{Roy, A.}
\newblock \bibinfo{title}{A practical heuristic for finding graph minors} (\bibinfo{year}{2014}).
\newblock \eprint{1406.2741}.

\bibitem{dwave-chainstrength}
\bibinfo{author}{{D-Wave Systems}}.
\newblock \bibinfo{title}{Programming the d-wave qpu: Setting the chain strength} (\bibinfo{year}{2020}).
\newblock \bibinfo{note}{White Paper}.

\bibitem{Greedy}
\bibinfo{author}{{D-Wave Systems}}.
\newblock \bibinfo{title}{dwave-greedy}.
\newblock \bibinfo{howpublished}{\url{https://github.com/dwavesystems/dwave-greedy}} (\bibinfo{year}{2022}).
\newblock \bibinfo{note}{Accessed: 2025-07-22}.

\bibitem{zucca2021diversitymetricevaluationquantum}
\bibinfo{author}{Zucca, A.}, \bibinfo{author}{Sadeghi, H.}, \bibinfo{author}{Mohseni, M.} \& \bibinfo{author}{Amin, M.~H.}
\newblock \bibinfo{title}{Diversity metric for evaluation of quantum annealing} (\bibinfo{year}{2021}).
\newblock \eprint{2110.10196}.

\end{thebibliography}

\section*{Acknowledgments}
This study was financially supported by programs for bridging the gap between R\&D and IDeal society (Society 5.0) and Generating Economic and social value (BRIDGE) and Cross-ministerial Strategic Innovation Promotion Program (SIP) from the Cabinet Office.
In addition, financial support was also provided through Sendai City's Advanced Technology and Data Utilization Use Case Creation Support Project.

\section*{Author contributions}
Matsumoto, Y., performed the experiments and analyzed the results.
Takabayashi, T., proposed problem formulation and created a calculation tool for problem solving.
Sato, R., and Honda, R., carried out the tasks necessary to conduct the experiment in Sendai City.
Ohzeki, M., conceived the idea of the analysis and supervised this work.
All authors participated in discussions of the results and contributed to the final manuscript.

\section*{Additional Information}
\subsection*{Data Availability}
The datasets used during the current study are available from the corresponding author upon reasonable request.

\end{document}